# ENHANCING FUZZ TESTING EFFICIENCY THROUGH AUTOMATED FUZZ TARGET GENERATION

**Tran Chi Thien**[a,*]

[a] *Faculty of Information Technology, Ton Duc Thang University*
*Ho Chi Minh City, Vietnam*
[*]*e-mail: tranchithien@tdtu.edu.vn*, ORCID: 0000-0003-3591-872X

Fuzzing continues to be the most effective method for identifying security vulnerabilities in software. In the context of fuzz testing, the fuzzer supplies varied inputs to fuzz targets, which are designed to comprehensively exercise critical sections of the client code. Various studies have focused on optimizing and developing advanced fuzzers, such as AFL++, libFuzzer, Honggfuzz, syzkaller, ISP-Fuzzer, which have substantially enhanced vulnerability detection in widely used software and libraries. Nevertheless, achieving greater coverage necessitates improvements in both the quality and quantity of fuzz targets. In large-scale software projects and libraries – characterized by numerous user defined functions and data types – manual creation of fuzz targets is both labor-intensive and time-consuming. This challenge underscores the need for automated techniques not only to generate fuzz targets but also to streamline the execution and analysis of their results. In this paper, we introduce an approach to improving fuzz target generation through static analysis of library source code. The proposed method encompasses several key aspects: it analyzes source code structures to accurately construct function calls and generate fuzz targets; it maps fuzzer input data to the corresponding function parameters; it synthesizes compilation information for the fuzz targets; and it automatically collects and analyzes execution results. Our findings are demonstrated through the application of this approach to the generation of fuzz targets for C/C++ libraries.

## 1. INTRODUCTION

Fuzz testing has experienced significant advancements in detecting security vulnerabilities in recent times [1]. It has identified thousands of bugs and vulnerabilities across a wide range of applications. In the context of fuzz testing, the fuzzer supplies varied inputs to fuzz targets, which are designed to comprehensively exercise critical sections of the client code. Expanding the range of sections covered directly improves the overall effectiveness of fuzz testing. U˘

Various studies have focused on optimizing and developing advanced fuzzers to enhance vulnerability detection in software systems. For instance, AFL++ is an evolution of the original American Fuzzy Lop (AFL) [2] and incorporates improved instrumentation, refined mutation strategies, and extended platform support to achieve higher code coverage and performance. libFuzzer [3], developed as part of the LLVM project, leverages in-process, coverage-guided fuzzing and integrates seamlessly with sanitizers to detect errors such as buffer overflows and memory corruptions. Honggfuzz [4] is another robust fuzzer that emphasizes flexibility and ease of use, offering features such as customized input mutations, crash reproduction, and adaptive heuristics suitable for a wide variety of target applications. In the realm of operating system kernels, syzkaller [5] has been specifically designed to uncover kernel-level vulnerabilities by systematically generating system calls and exploring the complex interfaces of kernel code.

Additionally, emerging fuzzers like ISP-Fuzzer [10] and Avalanche [8] introduce innovative approaches to further narrow the gap between the vast input space and the relatively sparse defect space. ISP-Fuzzer focuses on advanced input space partitioning techniques to optimize the fuzzing process for complex software systems, while Avalanche employs cutting-edge strategies to automatically generate fuzz targets, thereby overcoming some of the inherent limitations of traditional fuzzing methods.

Nevertheless, a study on Fuzz Blockers in [12] proves that achieving broader coverage requires enhancements in both the quality and quantity of fuzz targets. In large-scale software projects and libraries – characterized by a multitude of user-defined functions and data types – the manual creation of fuzz targets is both labor-intensive and time-consuming. This challenge highlights the necessity for automated techniques not only to generate fuzz targets but also to streamline their execution and result analysis.

In 2019, Google introduced Fudge [9], a system for automated fuzz target generation. Fudge leverages existing usage patterns of testing libraries to learn how to correctly invoke library functions. of In August 2023, Google launched LLM-aided fuzzing [13], an approach integrated with OSS-Fuzz [14] – a continuous fuzzing system introduced in 2016 – to assist developers in generating fuzz targets for their projects. This framework utilizes machine learning models and algorithms trained on programming languages and publicly available source code to generate test cases.

Google outlines the following steps in the code generation process using LLM-aided fuzzing:
1. Identification of under-fuzzed code: OSS-Fuzz's Fuzz Introspector tool detects high-potential, under-tested portions of the sample project's code and passes them to the evaluation framework.
2. Prompt generation: The evaluation framework constructs a prompt containing project-specific information, which is then used by the LLM to generate a new fuzz target.
3. Execution of the generated fuzz target: The evaluation framework compiles and executes

the newly generated fuzz target.
4. Coverage analysis: The evaluation framework monitors execution to assess any changes in code coverage.
5. Iterative refinement: If the generated fuzz target fails to compile, the evaluation framework prompts the LLM to revise the fuzz target, addressing compilation errors.

At first, the generated fuzz target wouldn't compile; however, after several rounds of prompt engineering and trying out the new fuzz targets, the code coverage of testing projects gain better.

Although the mentioned method finds many bugs in popular libraries, the LLM solution requires powerful computing and hardware capabilities. Additionally, without access to a large codebase (in the context of low-resource testing) – or if the testing libraries are newly developed and not widely used – fuzz target generation remains a challenging problem.

The research presented in this paper focuses on addressing the following issues:
1. Automation of static source code analysis to leverage dependencies between entities,
2. Generation of fuzz targets in both the presence and absence of consumer programs,
3. Feeding inputs to called functions in fuzz targets,
4. Automation of fuzzing and result analysis.

In 2020, the development of the Futag tool began at the Institute for System Programming of the Russian Academy of Sciences (ISP RAS) to address these issues. Futag provides solutions for the first and third challenges, with some details on the fuzz target generation process presented in [6] and [7]. It is capable of generating fuzz targets of libFuzzer and AFL++ for C/C++ libraries. However, a more detailed explanation of fuzz target generation methods is still necessary. This will be presented in the third section, while the fourth section will focus on the automation of fuzzing and result analysis.

## 2. FUZZING AND LIBRARY TESTING

To explore this topic in greater depth, it is essential to examine fuzzers, fuzz targets, and the process of fuzz testing for libraries.

In the diagram below (Fig. 1), a fuzzer supplies inputs to the fuzz target. These inputs may be generated dynamically or derived from an optional corpus provided by the user. The fuzz target then interacts with the testing library through its application programming interfaces (APIs). APIs abstract the complex logic of data manipulation within the library by receiving input data from the application or client code – in this case, from the fuzz target – and handling the creation of specific data structures. This key feature of APIs suggests an approach for generating user-defined arguments for function calls, which will be discussed in the next section.

For effective fuzz testing, the library must be compiled with sanitizers. During execution, the fuzzer monitors the behavior and code coverage of both the fuzz target and the library, facilitating the detection of potential vulnerabilities and unexpected behaviors.

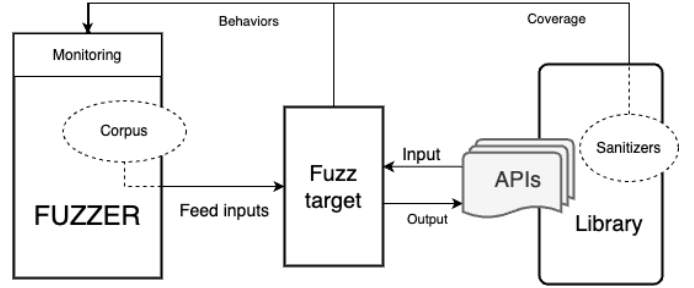

**Fig**. 1. Library fuzzing testing.

In [6] and [7] we have introduced method static analys for determining entities' relationships and extracting useful library APIs for fuzzing, the information is saved in knowledge database of testing library.

## 3. FUZZ TARGETS GENERATION

### 3.1. Generation without context of consumer programs

Let's start with function calls in C/C++. To execute a function correctly, the function call must match the function name exactly as declared in its prototype. The list of arguments must correspond precisely to the number of parameters required by the function. Additionally, each argument must have the same type as its corresponding parameter, and the return type in the function call must align with the return type specified in the function prototype.

If all parameters have fundamental or basic data types, we can easily initialize the arguments with inputs from the fuzzer.

However, the process becomes more complex when one or more parameters have user-defined types from the testing library. As discussed in Section 2, such parameters can be created using the library's APIs. The algorithm for generating function calls is illustrated in Figure 2. The input to GenFunctionCall is the prototype of the target function. Depending on the data type of each parameter, the appropriate method (e.g., GenPrimitiveValue, GenPointerValue, GenStructValue, etc.) is invoked to generate the corresponding argument for the function call. If a parameter's data type is a user-defined type from the testing library, the argument is initialized using an API call that returns the required data type.

**Algorithm 1**: Generating function calls without usage con-

```
texts
1   Function GenFunctionCall:
    Input  : function_info, library_info
    Ouput  : function_call
2   parameter_list ← GetParamList(function_info.prototype)
3   arg_values ← []
4   API_func_list ← GetAPIListFromDB()
5   FOR each param IN parameter_list DO:
6     IF param.type IS primitive THEN
7       value ← GenPrimitiveValue(param.type)
8     ELSE IF param.type IS pointer THEN
9       value ← GenPointerValue(param.base_type)
10    ELSE IF param.type IS struct THEN
11      value ← GenStructValue(param.fields)
12    ELSE IF param.type IS array THEN
13      value ← GenArrayValue(param.elem_type,
14  param.size)
15    ELSE
16      found_API ← FALSE
17      FOR each API_func IN API_func_list DO:
18        IF param.type IS API_func.return_type THEN
19          value ← GenFunctionCall (API_func.func_info)
20          found_API ← TRUE
21      IF found_API IS FALSE THEN
22        Return UNABLE_TO_GENERATE
23    arg_values.Append(value)
24  function_call ← ConstructFuncCall(arg_values)
25  RETURN function_call
```

```
1   FUNCTION GenFunctionCall(func_info):
2     parameter_list ← GetParamList(func_info.prototype)
3     arg_values ← []
4     API_func_list ← GetAPIListFromDB()
5     FOR each param IN parameter_list DO:
6       IF param.type IS primitive THEN
7         value ← GenPrimitiveValue(param.type)
8       ELSE IF param.type IS pointer THEN
9         value ← GenPointerValue(param.base_type)
10      ELSE IF param.type IS struct THEN
11        value ← GenStructValue(param.fields)
12      ELSE IF param.type IS array THEN
13        value ← GenArrayValue(param.elem_type,
    param.size)
14      ELSE
15        found_API ← FALSE
16        FOR each API_func IN API_func_list DO:
17          IF param.type IS API_func.return_type THEN
18            value ← GenFunctionCall (API_func.func_info)
19            found_API ← TRUE
20        IF found_API IS FALSE THEN
21          Return UNABLE_TO_GENERATE
22      arg_values.Append(value)
23    function_call ← ConstructFuncCall(arg_values)
24    RETURN function_call
```

**Fig**. 2. Generating function calls without usage contexts

As you may notice, when searching the knowledge database of the testing library, one or more results (APIs) may match the data type of the parameters. In such cases, the function GenFunctionCall is invoked recursively to generate all possible function calls. Consequently, there may be more than one valid function call for each function.

In object-oriented programming, a function may be a member method of a class. In this case, an instance of the class must be created first before generating the corresponding method call. The algorithm illustrating this process is shown below. The function GenMethodCall tries to initialize the instant of the class by various constructors, and then the initialized instant call its member method with help of GenFunctionCall.

```
1   FUNCTION GenMethodCall(method_info):
3     class_info = GetClassInfo(method_info)
4     constructor_list ← FindAllConstructors(class_info)
5     FOR each constr IN constructor_list DO:
6       instant ← GenClassConstructor (constr.prototype)
7       method_call = GenFunctionCall(instant. method_info)
8     RETURN method_call
```

**Fig**. 3. Generating method call

The realization of these algorithms can be found in an open Github repository [15].

The proposed method of fuzz target generation relies on knowledge of the testing library, which is gathered through static analysis during the compilation process [6]. It is important to emphasize that static analysis enables the exploration of all possible API call constructions that may be overlooked in the documentation. Additionally, static analysis can identify security vulnerabilities and critical errors in the source code [16].

### 3.2. Generation within contexts of consumer programs

The function call context of a library refers to the set of conditions and constraints under which its functions are invoked. This context encompasses various factors, including the state of the calling environment, dependencies between function calls, and the data structures involved in the execution process. As noted in Section 1, the LLM-aided fuzz target generation method leverages context from consumer programs within the codebase. We propose a fuzz target generation approach designed for low-resource cybersecurity environments that utilizes function contexts from consumer programs provided by the user.

To analyze a program's behavior during execution, both the control flow graph (CFG) [17] and the data flow graph (DFG) [18] are employed. The CFG illustrates the order in which statements, blocks, or functions are executed, highlighting branches, loops, and decision points. This visualization aids in examining potential execution paths and identifying unreachable code or possible errors. Conversely, the DFG maps the movement of data within the program by modeling the dependencies between variables and operations. This perspective is instrumental in detecting issues such as uninitialized variables, data races, and redundant computations.

In the proposed method, we analyze functions in the consumer program using both the Control Flow Graph (CFG) and the Data Flow Graph (DFG) to determine the usage contexts of the testing library. To utilize the library, the consumer program calls the library API, supplies input data, and assigns the return value to a variable (referred to as *init_var* in Figure 4). The consumer program may then invoke additional APIs before retrieving the processed output. The provided input is tracked and analyzed as it is manipulated by the testing library's APIs through various control flows in the consumer function. In other words, the data flow of the provided input from the consumer program to the testing library is evaluated for each control flow [19].

The algorithm of discovering usage context for fuzz target generating is described in Figure below.

```
1   FUNCTION DiscoverContext(lib_info, consumer_func):
2       all_contexts ← []
3       API_func_list ← GetAPIListFromDB ()
4       init_vars ← consumer_func.FindAllInitAPICalls()
5       CFG_paths ← consumer_func.FindAllCFG()
6       FOR each var IN init_vars DO:
7           FOR each CF_path IN CFG_paths DO:
8               context ← []
9               el_vars ← []
10              el_calls ← []
11              context. AppendVar(var)
12              FOR each instruction in CF_path:
13                  IF instruction IS PROCESS_BINARY:
14                      el_vars.add(instruction.left_path)
15                  IF instruction IS PROCESS_FUNC_CALL:
16                      el_calls.add(instruction.GetFuncCall())
17              context. AddVar(el_vars)
18              context. AddCall(el_calls)
19              all_contexts.add(context)
20      RETURN all_contexts
```

**Fig**. 4. Discovering usage contexts in consumer program

The data flow of a variable may consist of a list of derived variables – variables that originate from the analyzed variable – and a list of function calls that manipulate the analyzed variable. Following the study in [21], to evaluate the data flow of a variable (a usage context), the proposed method focuses on two main operations:

- Binary operations: If the evaluated variable appears on the right side of a binary operation (PROCESS_BINARY), the variable on the left side is added to the list of derived variables.
- Function calls: If the evaluated variable exists in the argument list of a function call (PROCESS_FUNC_CALL), that function call is added to the data flow.

With the discovered usage context, we can assign its variable with a buffer from the fuzzer [6] and construct API calls for the fuzz target (as discussed in Section 3.1). To evaluate the proposed method experimentally, we attempted to search for context and generate fuzz targets for the json-c library using libstorj as the consumer program. The steps of discovery are listed below.

- Compile the testing library to collect the knowledge database.
- Compile the consumer program/library to analyze its functions
- Analyze functions' consumer program to generate fuzz targets of testing library.

The Futag instrument successfully generated three fuzz targets with complex usage contexts (for more detail please visit [20]), such as:

- Creating a json_object *body* using the function *json_object_new_object*.
- Creating a string json_object named *FutagRefVar1U8* using the function *json_object_new_string*.
- Creating an int64 json_object named *FutagRefVarrfN* using the function *json_object_new_int64*.
- Creating an int json_object named *FutagRefVarwSm* using the function *json_object_new_int*.
- Creating an int64 json_object named *FutagRefVargXe* using the function *json_object_new_int64*.

- Calling the function *json_object_object_add* to add the object *FutagRefVar1U8* to *body* with the key "*token*".
- Calling the function *json_object_object_add* to add the object *FutagRefVarrfN* to *body* with the key "*exchangeStart*".
- Calling the function *json_object_object_add* to add the object *FutagRefVargXe* to *body* with the key "*exchangeEnd*".
- Calling the function *json_object_put*(*body*) to decrease the reference count of *body*.

The generated fuzz targets were successfully compiled and executed.

It is important to note that if the function under analysis is too complex, the number of control flows and data flows can become very large, significantly increasing the analysis time.

Table 1. Coverage of a single fuzz target on json-c library

| **Filename** | **Regions** | **Missed Regions** | **Cover** | **Functions** | **Missed Functions** | **Executed** |
|---|---|---|---|---|---|---|
| arraylist.c | 99 | 62 | 37.37% | 12 | 7 | 41.67% |
| json_object.c | 910 | 808 | 11.21% | 100 | 74 | 26.00% |
| json_tokener.c | 1104 | 441 | 60.05% | 16 | 6 | 62.50% |
| json_util.c | 97 | 75 | 22.68% | 13 | 11 | 15.38% |
| linkhash.c | 227 | 211 | 7.05% | 20 | 17 | 15.00% |
| linkhash.h | 10 | 10 | 0.00% | 8 | 8 | 0.00% |
| printbuf.c | 66 | 37 | 43.94% | 7 | 2 | 71.43% |
| random_seed.c | 41 | 41 | 0.00% | 4 | 4 | 0.00% |
| strerror_override.c | 27 | 27 | 0.00% | 1 | 1 | 0.00% |
| **TOTAL** | **2587** | **1712** | **33.82%** | **182** | **130** | **28.57%** |

Table 2. Overall coverage of all fuzz targets on json-c library

| **Filename** | **Regions** | **Missed Regions** | **Cover** | **Functions** | **Missed Functions** | **Executed** |
|---|---|---|---|---|---|---|
| arraylist.c | 99 | 61 | 38.38% | 12 | 6 | 50.00% |
| debug.c | 17 | 10 | 41.18% | 6 | 2 | 66.67% |
| json_object.c | 910 | 341 | 62.53% | 100 | 23 | 77.00% |
| json_object_iterator.c | 40 | 33 | 17.50% | 7 | 6 | 14.29% |
| json_pointer.c | 146 | 130 | 10.96% | 9 | 5 | 44.44% |
| json_tokener.c | 1104 | 370 | 66.49% | 16 | 4 | 75.00% |
| json_util.c | 97 | 15 | 84.54% | 13 | 1 | 92.31% |
| linkhash.c | 227 | 204 | 10.13% | 20 | 16 | 20.00% |
| linkhash.h | 10 | 9 | 10.00% | 8 | 7 | 12.50% |
| printbuf.c | 66 | 15 | 77.27% | 7 | 0 | 100.00% |
| random_seed.c | 41 | 41 | 0.00% | 4 | 4 | 0.00% |
| strerror_override.c | 27 | 20 | 25.93% | 1 | 0 | 100.00% |
| **TOTAL** | **2784** | **1249** | **55.14%** | **203** | **74** | **63.55%** |

Table 3. Found vulnarablities

| Library | Version | Function | Bug type | Date of report | Date of patch |
|---|---|---|---|---|---|
| libpng | 1.6.37 | png_convert_from_time_t | DEADLYSIGNAL | Feb 8, 2021 | Sep 13, 2022 |
| pugixml | 1.13 | default_allocate | allocation-size-too-big | Apr 11, 2023 | Apr 15, 2023 |

## 4. EVALUATION AND RESULTS

The proposed method is implemented using Futag, as described in [7]. The instrument Futag was executed on several popular libraries with the following hardware configuration:
- Platform: VMWare
- Number of processors: 4
- Number of cores per processor: 4
- Memory: 32GB
- Operating system: Ubuntu 22.04

The proposed method is evaluated by the quality of generated fuzz targets and the overall code coverage of their execution.

### 4.1. The quality of generated fuzz targets

The quality of generated fuzz targets is assessed based on the following criteria:
- Number of generated functions: The number of functions for which fuzz targets were generated.
- Number of compiled functions: The number of functions for which fuzz targets successfully compiled.
- Number of generated fuzz targets: The total number of fuzz targets generated.
- Number of compiled fuzz targets: The total number of fuzz targets that successfully compiled.

The result of generation is shown in Table 4.

Futag can generate dozens (even hundreds) of fuzz targets in just a few minutes, with the majority successfully compiling and executing. As a result, the workload for fuzz target generation is significantly reduced.

Table 4. Generated fuzz targets for popular libraries

| Library | № generated functions | № compiled functions | № generated fuzz targets | № compiled fuzz targets |
|---|---|---|---|---|
| **json-c** | 134 | 84 | 539 | 448 |
| **pugixml** | 121 | 94 | 201 | 151 |
| **tinyxml** | 63 | 25 | 77 | 39 |

Below are some cases in which Futag cannot generate fuzz targets or where the parameters of the function call in the generated fuzz target are not correctly initialized:
1. There is insufficient information about the data type for generation. For example, a function argument has an integer type that represents a file descriptor. The file descriptor value is assigned after creating a new input-output stream; therefore, if a mutated value is passed, the generation is incorrect. Similarly, if a function argument has a string type that represents a file path, the string must be passed in a system path format; otherwise, the generated result will be incorrect.
2. Generating a fuzzing wrapper for class constructors that have a parameter of a derived complex type.
3. A function argument has a data type that is not currently supported for analysis.
4. A function argument has a complex derived type and is initialized using a function that accepts a complex type as an argument.

In cases where fuzz targets fail to compile, static analysis identifies missing compilation information, such as included headers or required libraries during the linking stage.

### 4.2. The overall coverage

To gather information on overall coverage, we execute each generated fuzz target for approximately 10 seconds and use the LLVM-Cov tool for analysis. LLVM-Cov computes the code coverage for each fuzz target individually and then merges the results to determine the overall code coverage of the testing library.

The code coverage of single fuzz target is shown in Table 1, and the overall code coverage achieved using the proposed method for testing the json-c library is presented in Table 2.

It is important to note that the fuzz targets are executed without a corpus, which plays a crucial role in fuzzing structured data, such as JSON files. The results indicate that while the coverage of a single fuzz target may not be extensive, the combined execution of multiple fuzz targets achieves good overall coverage.

*4.3. Discovered Vulnerabilities*

The instrument has identified some vulnerabilities in popular libraries such as libpng and pugixml (Table 3).

## 5. CONCLUSIONS

The proposed method generates fuzz targets for libraries—both with and without usage contexts—while requiring minimal resources. Moreover, the developed tool is capable of detecting new errors in popular libraries. The proposed method relies on the results of static analysis; however, to improve the quality of the generated fuzz targets and address the issues discussed in Section 4.1, a more in-depth analysis is needed to better understand the relationships between entities.

The number of fuzz targets generated from the consumer program depends on the number of usage contexts and the complexity of the consumer program. Although the method currently analyzes two main types of instructions, enhancing the approach will require analyzing more complex instructions, such as loops and conditional statements.

Further research can proceed for other programming languages like python, JAVA, etc.

## REFERENCES

[1]. *Böhme M, Cadar C, Roychoudhury A.* Fuzzing: Challenges and reflections. IEEE Software. 2020 Aug 13;38(3):79-86.

[2]. *Zalewski M.* Afl: American fuzzy lop. URL: https://github.com/mirrorer/afl. 2021 Jan.

[3]. *Serebryany K.* Continuous fuzzing with libfuzzer and addresssanitizer. In 2016 IEEE Cybersecurity Development (SecDev) 2016 Nov 3 (pp. 157-157). IEEE.

[4]. *Google.* Syzkaller. https://github.com/google/syzkaller.

[5]. *Google.* Honggfuzz. https://github.com/google/honggfuzz

[6]. *C. T. Tran and S. Kurmangaleev*, "Futag: Automated fuzz target generator for testing software libraries," 2021 Ivannikov Memorial Workshop (IVMEM), Nizhny Novgorod, Russian Federation, 2021, pp. 80-85, doi: 10.1109/IVMEM53963.2021.00021.

[7]. *C. T. Tran, D. Ponomarev and A. Kuznhesov*, "Research on automatic generation of fuzz-target for software library functions," *2022 Ivannikov Ispras Open Conference (ISPRAS)*, Moscow, Russian Federation, 2022, pp. 95-99, doi: 10.1109/ISPRAS57371.2022.10076871.

[8]. *Ermakov M.K., Gerasimov A.Y.* Avalanche: adaptation of parallel and distributed computing for dynamic analysis to improve performance of defect detection. *Proceedings of the Institute for System Programming of the RAS (Proceedings of ISP RAS)*. 2013;25:29-38. (In Russ.)

[9]. *Babić D, Bucur S, Chen Y, Ivančić F, King T, Kusano M, Lemieux C, Szekeres L, Wang W.* Fudge: fuzz driver generation at scale. In Proceedings of the 2019 27th ACM Joint Meeting on European Software Engineering Conference and Symposium on the Foundations of Software Engineering 2019 Aug 12 (pp. 975-985).

[10]. *Sargsyan S, Hakobyan J, Mehrabyan M, Mishechkin M, Akozin V, Kurmangaleev S.* ISP-Fuzzer: Extendable fuzzing framework. In 2019 Ivannikov Memorial Workshop (IVMEM) 2019 Sep 13 (pp. 68-71). IEEE.

[11]. *Ispoglou, Kyriakos, Daniel Austin, Vishwath Mohan, and Mathias Payer.* "{FuzzGen}: Automatic fuzzer generation." In *29th USENIX Security Symposium (USENIX Security 20)*, pp. 2271-2287. 2020.

[12]. *Wentao Gao, Van-Thuan Pham, Dongge Liu, Oliver Chang, Toby Murray, and Benjamin I.P. Rubinstein.* 2023. Beyond the Coverage Plateau: A Comprehensive Study of Fuzz Blockers (Registered Report). In Proceedings of the 2nd International Fuzzing Workshop (FUZZING 2023). Association for Computing Machinery, New York, NY, USA, 47–55, doi: 10.1145/3605157.3605177

[13]. *Google.* AI-Powered Fuzzing: Breaking the Bug Hunting Barrier, https://security.googleblog.com/2023/08/ai-powered-fuzzing-breaking-bug-hunting.html

[14]. *Serebryany, Kostya.* "{OSS-Fuzz}-Google's continuous fuzzing service for open source software." (2017).

[15]. *Futag.* https://github.com/ispras/Futag/

[16]. *Avetisyan A, Belevantsev A, Borodin A, Nesov V.* Using static analysis for finding security vulnerabilities and critical errors in source code. Proceedings of the Institute for System Programming of the RAS (Proceedings of ISP RAS). 2011;21.

[17]. *Allen, F. E.* Control Flow Analysis. Proceedings of a Symposium on Compiler Optimization. — Urbana-Champaign, Illinois: Association for Computing Machinery, 1970. — C. 1—19, doi: 10.1145/800028.808479.

[18]. *Kennedy K.* A survey of data flow analysis techniques. IBM Thomas J. Watson Research Division; 1979.

[19]. *Allen FE, Cocke J.* A program data flow analysis procedure. Communications of the ACM. 1976 Mar 1;19(3):137.

[20]. *Futag-test.* Futag-tests/json-c-contexts/succeeded/json_object_put/json_object_put.1

[21]. *Weiser M.* Program slicing. IEEE Transactions on software engineering. 2009 May 29(4):352-7.